\documentclass[a4paper,11pt]{article}
\usepackage[top=3.25cm,bottom=3.25cm,left=2.75cm,right=2.75cm,a4paper]{geometry}
\usepackage{amsfonts,amsmath,amssymb,amsthm}
\usepackage{float,xcolor,fancyhdr,siunitx}
\usepackage{booktabs,longtable,tabularx}
\usepackage{setspace}
\usepackage{lineno}
\usepackage[normalem]{ulem} 
\usepackage[colorlinks]{hyperref}
\hypersetup{
    colorlinks=true, 
    linkcolor={blue!50!black},
    citecolor={green!60!black},
    urlcolor= {red!80!black}
}
\usepackage{enumitem,mdframed}
\usepackage{empheq}
\usepackage{upgreek}
\usepackage{multirow}
\usepackage[]{algorithm2e}
\usepackage[bottom]{footmisc} 
\usepackage{tikz,pgfplots,pgfplotstable}
\usepgfplotslibrary{groupplots}
\usetikzlibrary{positioning,arrows,fit,patterns}
\usetikzlibrary{backgrounds}
\pgfplotsset{compat=newest}
\usepackage{arydshln}

\usepackage[labelformat=simple]{subcaption}       
\DeclareCaptionLabelFormat{opening}{(#2)} 
\makeatletter
\renewcommand*\env@matrix[1][\arraystretch]{%
  \edef\arraystretch{#1}%
  \hskip -\arraycolsep
  \let\@ifnextchar\new@ifnextchar
  \array{*\c@MaxMatrixCols c}}
\makeatother

\usepackage[]{todonotes}

\newcommand{\RR}{\mathbb{R}}

\newcommand{\flpr}{\mathrm{p}}

\newcommand{\Rsun}{R_\odot}
\newcommand{\ra}{r_a}
\newcommand{\rs}{r_{\mathrm{s}}}
\newcommand{\vel}{\mathrm{c}_0}
\newcommand{\bx}{\mathbf{x}}
\newcommand{\density}{\rho_0}
\newcommand{\Ehe}{E_{\mathrm{he}}}
\newcommand{\tEhe}{\tilde{E}_{\mathrm{he}}}

\newcommand{\atmogamma}{\gamma_{\mathrm{a}}}
\newcommand{\atmovel}  {\mathrm{c}_{\mathrm{a}}}
\newcommand{\atmoalpha}{\alpha_{\mathrm{a}}}

\newcommand{\gammaspline} {\gamma_{\mathrm{s}}}
\newcommand{\logrhospline}{\varrho_{\mathrm{s}}}

\newtheorem{rmk}{Remark}

\newlength{\plotwidth}  
\newlength{\plotheight} 
\newcommand{\datafile}{}

\newcommand{\dataA}{} \newcommand{\legendA}{}

\newcommand{\myylabel} {}
\newcommand{\legendpos}{}
\newcommand{\mlink}{\url{http://phaidra.univie.ac.at/o:1097638}}
\newcommand{\zxmin}{}
\newcommand{\zxmax}{}
\newcommand{\zxminbox}{}
\newcommand{\zxmaxbox}{}
\newcommand{\zymin}{}
\newcommand{\zymax}{}
\newcommand{\xticka}{}
\newcommand{\xtickb}{}

\usepackage[capitalize,nameinlink]{cleveref}
\crefname{section}   {Section}   {Sections}
\crefname{subsection}{Subsection}{Subsections}
\Crefname{section}   {Section}   {Sections}
\Crefname{subsection}{Subsection}{Subsections}
\Crefname{figure}    {Figure}    {Figures}

\crefformat{equation}{\textup{#2(#1)#3}}
\crefrangeformat{equation}{\textup{#3(#1)#4--#5(#2)#6}}
\crefmultiformat{equation}{\textup{#2(#1)#3}}{ and \textup{#2(#1)#3}}
{, \textup{#2(#1)#3}}{, and \textup{#2(#1)#3}}
\crefrangemultiformat{equation}{\textup{#3(#1)#4--#5(#2)#6}}%
{ and \textup{#3(#1)#4--#5(#2)#6}}{, \textup{#3(#1)#4--#5(#2)#6}}{, and \textup{#3(#1)#4--#5(#2)#6}}
\Crefformat{equation}{#2Equation~\textup{(#1)}#3}
\Crefrangeformat{equation}{Equations~\textup{#3(#1)#4--#5(#2)#6}}
\Crefmultiformat{equation}{Equations~\textup{#2(#1)#3}}{ and \textup{#2(#1)#3}}
{, \textup{#2(#1)#3}}{, and \textup{#2(#1)#3}}
\Crefrangemultiformat{equation}{Equations~\textup{#3(#1)#4--#5(#2)#6}}%
{ and \textup{#3(#1)#4--#5(#2)#6}}{, \textup{#3(#1)#4--#5(#2)#6}}{, and \textup{#3(#1)#4--#5(#2)#6}}

\crefdefaultlabelformat{#2\textup{#1}#3}
\crefname{prop}{Proposition}{Propositions}
\Crefname{prop}{Proposition}{Propositions}
\crefname{definition}{Definition}{Definitions}
\Crefname{definition}{Definition}{Definitions}
\crefname{thm}{Theorem}{Theorems}
\Crefname{thm}{Theorem}{Theorems}
\crefname{rmk}{Remark}{Remarks}
\Crefname{rmk}{Remark}{Remarks}
\crefname{lem}{Lemma}{Lemma}
\Crefname{lem}{Lemma}{Lemma}
\crefname{assumption}{Assumption}{Assumptions}
\Crefname{assumption}{Assumption}{Assumptions}

\title{
$\mathcal{C}^2$ representations of the solar background 
coefficients for the model $\texttt{S-AtmoI}$.}

\author{
Florian Faucher\thanks{Faculty of Mathematics, University of Vienna, Oskar-Morgenstern-Platz 1,
                       A-1090 Vienna, Austria.
                      (\href{mailto:florian.faucher@univie.ac.at}
                      {\texttt{florian.faucher@univie.ac.at}}).                   
                    }
\and
Damien Fournier\thanks{Max-Planck-Institut f\"ur 
                       Sonnensystemforschung, Justus-von-Liebig-Weg 3, 
                       37077 G\"ottingen, Germany.}
\and
Ha Pham\thanks{Inria Project-Team Magique 3D, E2S--UPPA, CNRS, Pau, France.}}

\date{\today}

\begin{document}
\maketitle 

\numberwithin{equation}{section}

\begin{abstract}

  We construct $\mathcal{C}^2$ representations of 
  the background quantities that characterize the 
  interior of the Sun and its atmosphere starting 
  from the data-points of the standard solar 
  model~\texttt{S} of \cite{christensen1996current}. 
  This model is further extended considering an 
  isothermal atmosphere, that we refer to as \emph{model~\texttt{AtmoI}}. 
  It is not trivial to build the $\mathcal{C}^2$ representations 
  of the parameters from a discrete set of values, in particular
  in the transition region between the end of model \texttt{S}
  and the atmosphere. 
  This technical work is needed as a crucial building block to 
  study theoretically and numerically the propagation of waves in the Sun, 
  using the equations of solar oscillations (also referred to as
  Galbrun's equation in aeroacoustics).
  The constructed models are available at \mlink.

\end{abstract}


\section{Introduction}

In this work, we construct a $\mathcal{C}^2$ representation 
for the spherically symmetric background parameters 
characterizing the Sun, based on the model \texttt{S} for 
the interior, combined with an isothermal atmospheric model 
denoted \texttt{AtmoI}.
The propagation in the Sun is given by the vectorial Galbrun's 
equation, which describes the adiabatic wave motion on top of a 
static fluid background at equilibrium, and is characterized by
the following medium properties:
\begin{equation}\label{level0par}
  \text{the density } \uprho_0 \, , \qquad
  \text{the adiabatic index } \upgamma \, , \qquad
  \text{the fluid pressure } p_0 \, ,
\end{equation}
and the following auxiliary parameters: 
\begin{equation}\label{level1par}
  \text{the adiabatic sound speed } c_0 \qquad \text{and} \qquad
  \text{the gravitational potential } \phi_0\,.
\end{equation}
By \emph{auxiliary} quantities \cref{level1par}, we mean that they 
are derived from the principal parameters \cref{level0par}.
The principal parameters \cref{level0par} in the solar interior
are given by model \texttt{S}, however, under hydrostatic equilibrium, 
we obtain the pressure from $\uprho_0$ and $\upgamma$, and only use 
the value of the pressure given in the last entry of model \texttt{S}, 
see \cref{rep_p::sec}.
We refer to  \cite{barucq:hal-02423882} for a discussion 
of a simplified version of Galbrun's equation without flow, 
rotation and gravity potential perturbation.

The model \texttt{S} given in \cite{christensen1996current}
provides a point-wise representation of the principal parameters 
of \cref{level0par} up to a few hundred kilometers above the solar 
surface, but it is not satisfactory as the derivatives of the 
background parameters also appear in the wave equation, thus 
requiring to build $\mathcal{C}^2$ representations.
Namely, the Galbrun's vector equation in \cite{barucq:hal-02423882} 
requires the derivatives up to the second order of all of the 
physical parameters, and up to the third order for $p_0$.
In the scalar case that is obtained from the Galbrun's equation 
under simplifying assumptions, and that is mostly used in recent
works 
\cite{gizon2017computational,barucq2018atmospheric,fournier2017atmospheric,barucq:hal-02168467,OutEsaim},
the equation depends on $c_0$, $\uprho_0$ and the derivatives 
of $\uprho_0$ up to its second order.

In addition to building $\mathcal{C}^2$ representations, 
we extend the solar background quantities \cref{level0par} 
given by the model~\texttt{S} \cite{christensen1996current} 
beyond the surface of the Sun to take into account the 
presence of an atmosphere.
In our work, we introduce the isothermal atmospheric model~\texttt{AtmoI}, 
which offers one option to generalize the model \texttt{Atmo} employed in 
in the scalar case in \cite{fournier2017atmospheric,barucq2018atmospheric,barucq:hal-02168467,OutEsaim}.
Our model~\texttt{AtmoI} retains the exponential decay of $\uprho_0$ 
and constant $c_0$ of model~\texttt{Atmo}, however it needs additional 
assumptions to acknowledge the vector equation, see~\cite{barucq:hal-02423882}.

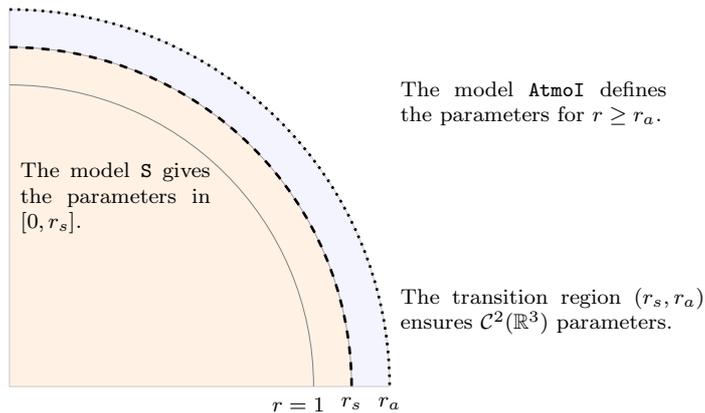
\begin{figure}[ht!] \centering
\begin{tikzpicture}[scale=1.,every node/.append style={font=\scriptsize}]
  \pgfmathsetmacro{\radiusMAX}{5.}
  \coordinate (x0) at (0,0) ;
  \pgfmathsetmacro{\rsurf}{0.80*\radiusMAX}
  \pgfmathsetmacro{\rs}   {0.90*\radiusMAX}
  \pgfmathsetmacro{\ra}   {1.00*\radiusMAX}
  \coordinate (x01) at (0,\rsurf) ;
  \coordinate (x02) at (\rsurf,0) ;
  \coordinate (xs1) at (0,\rs) ;
  \coordinate (xs2) at (\rs,0) ;
  \coordinate (xa1) at (0,\ra) ;
  \coordinate (xa2) at (\ra,0) ;  
  \draw[line width=0,fill=orange!50!white,opacity=0.2]  (x0) -- (xs1) to[in=90,out=0] (xs2) -- (x0);
  \draw[line width=0,fill=blue!20!white,opacity=0.2]    (xs1)-- (xa1) to[in=90,out=0] (xa2) -- (xs2) to[in=0,out=90] (xs1);
  \draw[line width=0.25,gray]    (x01) to[in=90,out=0] (x02);
  \draw[line width=1,dashed]  (xs1) to[in=90,out=0] (xs2);
  \draw[line width=1,dotted]  (xa1) to[in=90,out=0] (xa2);
    
  \node[line width=1,anchor= north, xshift=-2mm] at (x02) {$r=1$};  
  \node[line width=1,anchor= north] at (xs2) {$r_s$};
  \node[line width=1,anchor= north] at (xa2) {$r_a$};  
  \node[line width=1,text width=2.5cm,anchor=west,align=justify] at (0,\radiusMAX/2)
                                     {The model \texttt{S} gives 
                                      the parameters in $[0,r_s]$.};
  \node[line width=1,text width=3.5cm,anchor=west,align=justify] at (\radiusMAX,\radiusMAX*75/100) 
                  {The model \texttt{AtmoI} defines 
                   the parameters for $r \geq r_a$.};
  \node[line width=1,text width=4cm,anchor=west,align=justify] at (\radiusMAX,\radiusMAX*20/100) 
                  {The transition region $(r_s, r_a)$
                   ensures $\mathcal{C}^2(\mathbb{R}^3)$ parameters.};

\end{tikzpicture}
\caption{Illustration of the context: the model \texttt{S} of \cite{christensen1996current} gives
         point-wise coefficients for the (spherical) solar parameters $\uprho_0$, $\upgamma$ and 
         $p_0$ up to the position $r_s > 1$. We impose our atmospheric model \texttt{AtmoI} after
         $r_a > r_s$. The transition region $(r_s,r_a)$ is required to ensure globally $\mathcal{C}^2$
         coefficients. Here, $r$ denotes the scaled radius (see \cref{section:notation}), with 
         $r=1$ corresponding to the solar surface.}
\label{figure:illustration}
\end{figure}

By $\mathcal{C}^2$ representations, we mean the construction 
of $\mathcal{C}^2([0,\infty))$ functions for each of the 
quantities in \cref{level0par,level1par}. They coincide with 
the data points of the model \texttt{S} in the interior of the 
Sun (and satisfy the hydrostatic equilibrium) and the assumptions 
of the model~\texttt{AtmoI} in the atmosphere, see discussion 
in \cref{Satmocai::sec}. 
For these functions to remain globally $\mathcal{C}^2$, we 
need a transition region between the interior of the Sun and 
the atmosphere model as illustrated in \cref{figure:illustration}.

After introducing the notations of background parameters in 
scaled coordinates in \cref{section:notation}, we review the 
assumptions of model \texttt{S-AtmoI} in \cref{Satmocai::sec} and 
construct the background functions, with the following hierarchy 
of dependence:
\begin{itemize}
  \item[$\bullet$] We first construct $\gamma$ and $\uprho_0$, 
                   first in the interior (\cref{section:spline-modelS}) 
                   and then in the transition and the atmosphere (\cref{rep_gamma::sec}).
  \item[$\bullet$] After this step, the pressure $p_0$ and the gravitational 
                   potential $\phi_0$ and its first and second-order derivatives 
                   are obtained everywhere, see~\cref{rep_p::sec}.
  \item[$\bullet$] In the last step, we compute the remaining auxiliary quantities.
\end{itemize}
The constructed representations are made available at \mlink, together
with the script to reproduce the models and to obtain their values at 
any positions. 

\section{Notation and properties of model \texttt{S-AtmoI}}

In this section, we introduce the scaled radius and define 
the background coefficients in this new variable. We also 
specify the properties of the model~\texttt{S-AtmoI} and 
summarize the quantities in \cref{table:properties}.
All quantities are assumed to be spherically symmetric,
that is, to only depend on the position along the Sun's radius.

\subsection{Physical parameters and scaled variables}
\label{section:notation}

\paragraph{Physical constants} We use the following notation.
\begin{itemize}
\item[$\bullet$] We denote by  $\Rsun$ the Sun's radius with $\Rsun = \num{696.e8}$ \si{\cm}.
\item[$\bullet$] $G$ is the gravitational constant with $G=\num{6.67430e-8}$ \si{\cm\cubed\per\gram\per\second\squared}.
\end{itemize}
 
\paragraph{Scaled radius}
We denote by $\check{\bx}$ the 3D coordinate system 
with its origin at the center of the Sun, where the 
surface of the Sun is represented by $\{\check{\mathbf{x}}  \, | \,\,\,\, \lvert \check{\mathbf{x}}\rvert = \Rsun \}$.  
The scaled coordinates  $\bx \in \mathbb{R}^3$ and radius 
$r = \lvert \bx\rvert$,
are defined by
\begin{equation}
 \bx \, = \, \dfrac{\check{\bx}}{\Rsun} \, , \qquad\qquad r = \dfrac{R}{\Rsun} \,,
\end{equation}
where $R =\lvert \check{\bx}\rvert$. 
We note three special values of the scaled radius
\begin{equation}
   r= 1 \quad <\quad \rs \quad < \quad \ra\,.
\end{equation} 
The solar surface is located at $r=1$, $\rs$ 
is the last point of the model~\texttt{S}, and 
$\ra$ is the position where the solar atmosphere 
begins, see \cref{figure:illustration}. We further 
note that
\begin{itemize} 
  \item[$\bullet$] $\rs = \num{1.000716}$, corresponding to a height 
                   of about $\num{496}$ \si{\km} above the solar surface. 
  \item[$\bullet$] We choose $\ra = \num{1.000730}$, corresponding to $508$ \si{\km} above the surface. 
  \item[$\bullet$] The interval $(\rs, \ra)$ is referred to as the \emph{transition region}.
\end{itemize}

\paragraph{Original background parameters}

The original parameters are functions of the 
unscaled radius $R$, we have 
\begin{itemize}
  \item the density $\uprho_0(R)$ given in \si{\g\per\cm\cubed},
  \item the adiabatic index $\upgamma(R)$,
  \item the pressure $p_0(R)$ given in \si{\g\per \second\squared \per\cm}.
\end{itemize}
From these quantities, the \emph{(adiabatic) sound 
speed} (in \si{\cm\per\second}) is defined as
\begin{equation}
  c_0(R) \,\, := \,\, \sqrt{\dfrac{ \upgamma(R)\, p_0(R)  }{  \uprho_0(R)}}\,.
\end{equation}
The \emph{gravitational potential} $\phi_0(R)$ is the solution to 
\begin{equation}\label{original_phi0::def}
  \Delta_{\mathbf{x}} \phi_0 \, = \, 4\pi \, G \, \uprho_0
  \qquad\quad \text{given by} 
  \qquad\phi_0(\bx) \, := \, - G \int_{\RR^3} \dfrac{\density(\lvert \check{\mathbf{y}}\rvert)}
            {\lvert \check{\bx} \, - \, \check{\mathbf{y}}\rvert} 
  \, d \mathbf{y} \,.
\end{equation}
 
\paragraph{Backgrounds parameters in scaled radius}

From a function $f$ given in terms of $R$ to define a 
function $\mathsf{f}$ in terms of $r$, we use the construction, 
\begin{equation}\label{scaledfcn::def}
  r \,  \mapsto\,\, \mathsf{f}(r)\,  =\,  f(\Rsun\,  r)\, , \quad \text{for a given function  } f: R\,\mapsto\, f(R)\,.
\end{equation}
\begin{itemize}
  \item[$\bullet$] From $R\mapsto \uprho_0(R)$ and $R\mapsto \upgamma(R)$, we define respectively, 
  \begin{equation}
    \text{the density } \rho_0(r) \,\qquad (\text{in } \si{\g\per\cm\cubed})\qquad \text{and } 
    \qquad  \text{the adiabatic index } \gamma(r)\,.
  \end{equation}

 \item[$\bullet$] The scaled pressure field $\flpr_0(r)$ (in \si{\g\per\second\squared\per\cm\cubed}) is defined by
                 \begin{equation}
                   \flpr_0(r) \, = \, \dfrac{ p_0( \Rsun r)}{\Rsun^2}\,.
                 \end{equation}

 \item[$\bullet$] The scaled adiabatic background sound speed 
                 (given in \si{\per\second}) is defined as
                  the ratio 
                \begin{equation}\label{vel::def}
                  \vel(r) \, := \, \sqrt{\dfrac{\gamma(r)\, \flpr_0(r) }{\density(r)}} \, .
                \end{equation}
                
 \item[$\bullet$] The scaled background gravitational potential $\Phi_0(r)$ is 
                 \begin{equation}
                 \mathbf{x} \mapsto \Phi_0(\mathbf{x}) 
                 \, = \, \dfrac{ \phi_0(\Rsun\, \mathbf{x})}{\Rsun^2 }\,.
                 \end{equation}                            
                 It takes unit in $ \si{\per\second\squared}$ and is the solution to
                 \begin{equation}\label{Phi0::def}
                 \Delta_{\mathbf{x}} \Phi_0 \, = \, 4\pi G \rho_0
                 \quad \text{ given by } \quad   \Phi_0(\bx) \, := \, - G \int_{\RR^3} 
                 \dfrac{\density(\lvert \mathbf{y}\rvert)}{\lvert \bx - \mathbf{y}\rvert} 
                 \, d \mathbf{y} \,.
                 \end{equation}
\end{itemize}

\paragraph{Scale height functions}

The inverse scale height functions for a 
scalar $\mathcal{C}^1$ function $r\mapsto \mathfrak{g}(r)$ 
is defined as 
\begin{equation}\label{scale_heights::def}
  \alpha_{\mathfrak{g}}(r) \, := \, -\dfrac{\partial_r \mathfrak{g}(r)}{\mathfrak{g}(r)}
                           \,  = \, -\dfrac{\mathfrak{g}'(r)}{\mathfrak{g}(r)} \, .
\end{equation}

\paragraph{Hydrostatic equilibrium function} 
  Under the radial and regularity assumptions, 
  we work with the quantity $\Ehe$ and $\tEhe$, 
  that capture the hydrostatic equilibrium 
  condition, 
 \begin{equation}\label{Ehe::def}
    \Ehe \,\,:= \,\, \dfrac{ \Phi'_0}{\vel^2}\,\, \,-\,\dfrac{\alpha_{\flpr_{0}}}{\gamma} 
    \,, \qquad \quad \tEhe \,\,:= \,\, \rho_0 \, \Phi_0'\, + \, \flpr_0'  \,.
 \end{equation}

\subsection{Properties of model \texttt{S} and \texttt{AtmoI}}
\label{Satmocai::sec}

We will state the equations for the scaled parameters,
that is, with respect to the scaled radius $r$.

\paragraph{Model \texttt{S}} 
In this model, which is applied in the region $0\leq r \leq \rs$, 
the physical quantities $\rho_0$, $\flpr_0$, $\gamma$ are in 
hydrostatic equilibrium
\begin{equation}\label{hydrostatic}
   \flpr_0' \, =\, - \rho_0 \, \Phi_0'\quad  \text{on} \quad  r \in  [0,\rs]\, \qquad
   \text{hydrostatic equilibrium.}
\end{equation}
The above property comes from the Euler's equation under 
adiabatic assumption and with exterior force given 
by $\,-\rho_0 \nabla \Phi_0$, 
cf. \cite[Eqs (1.14--1.17) p.22]{legendre2003rayonnement}, 
and without flow. 
The Euler system reduces to the equation of motion which 
takes the form $\nabla \flpr_0  \, = \, -\rho_0 \nabla \Phi_0$. 
The hydrostatic relation \cref{hydrostatic} is its form 
in spherical symmetry.
Property \cref{hydrostatic} implies that 
\begin{equation}
   \Ehe \, \,\equiv\,\, 0 \,,\qquad \tEhe \, \,\equiv\,\, 0 \,, \qquad\text{ for }  0\leq r\leq \rs\,.
\end{equation}

\paragraph{Model \texttt{AtmoI}} In this model which is applied 
for the region $r\geq \ra$ that represents the extended atmosphere, 
the physical parameters $\rho_0$, $\vel$ and $\gamma$ satisfy
\begin{enumerate}
 \item the sound speed $\vel$ is constant and is equal to $\atmovel$, 
 \item the adiabatic coefficient $\gamma$ is constant and is equal to 
 $\atmogamma$,
 \item $\rho_0$ is exponentially decreasing, which implies 
       that the inverse density scale height $\alpha_{\rho_0}$ 
       is constant and is equal to $\atmoalpha$.
\end{enumerate}
The fluid pressure $\flpr_0$ follows from \cref{vel::def} and is given by 
$  \flpr_0 \, = \, {\vel^2\, \rho_0}/{\gamma}\,$.

\begin{table}[ht!] \begin{center}
\caption{Summary of the background quantities and properties, 
         defined in terms of the scaled radius $r$, and definition 
         of the model \texttt{AtmoI} in the atmosphere ($r \geq \ra$).}
\label{table:properties}
\renewcommand{\arraystretch}{1.20}
\begin{tabular}{|>{\centering\arraybackslash}p{.04\linewidth}|
                 >{\arraybackslash}p{.54\linewidth}|
                 >{\arraybackslash}p{.33\linewidth}|}
\hline
$\rho_0$ & density (in \si{\g\per\cm\cubed}) & principal parameter \\ \hline
$\gamma$ & adiabatic index                   & principal parameter \\ \hline
$\flpr_0$& fluid pressure (\si{\g\per \second\squared\per\cm\cubed})
         & principal parameter \\ \hline
$\alpha_\mathfrak{g}$ & scale height function of $\mathfrak{g}$  & $\alpha_\mathfrak{g} \,=\, -\mathfrak{g}' / \mathfrak{g}$. \\ \hline
  & & \\[-1.25em]
$\vel$   & scaled velocity (in \si{\per\second}) 
           from adiabaticity
         & $\vel(r) \, := \, \sqrt{\dfrac{\gamma(r)\, \flpr_0(r) }{\density(r)}}$
         \\ \hline
  & & \\[-1.25em]
$\Phi_0$ & scaled background gravitational potential (in \si{\per\second\squared})
         & $\Phi_0(\bx) \, := \, - G \int_{\RR^3} 
            \dfrac{\density(\lvert \mathbf{y}\rvert)}{\lvert \bx - \mathbf{y}\rvert} 
            \, d \mathbf{y} \,$ \\[0.80em] \hline
  & & \\[-1.25em]
$\Ehe$   & hydrostatic equilibrium representative function &
           $\Ehe \, := \, \dfrac{\Phi'_0}{\vel^2}\,\, \,-\,\dfrac{\alpha_{\flpr_{0}}}{\gamma}$ 
         \\[0.80em] \hline
$\tEhe$  & hydrostatic equilibrium representative function &
           $\tEhe \, := \, \rho_0 \, \Phi_0'\, + \, \flpr_0'$ 
         \\ \hline \multicolumn{3}{|c|}{} \\[-1.00em] \hline
         & hydrostatic equilibrium property & 
         $\flpr_0'=-\rho_0\Phi_0' \Rightarrow \Ehe=\tEhe=0$
         \\ \hline \multicolumn{3}{|c|}{} \\[-1.00em] \hline
  & \multicolumn{2}{c|}{} \\[-1.25em]
         & \multicolumn{2}{c|}{In the model \texttt{AtmoI}, for $r \geq \ra$,
                               we have:} \\
         & \multicolumn{2}{l|}{\hspace{1cm} constant velocity: $\vel \, =\, \atmovel$} \\
         & \multicolumn{2}{l|}{\hspace{1cm} constant adiabatic index $\gamma \, =\, \atmogamma$} \\
         & \multicolumn{2}{l|}{\hspace{1cm} exponentially decreasing $\rho_0$, i.e., 
                                            constant $\alpha_{\rho_0}=\atmoalpha$} \\
         & \multicolumn{2}{l|}{\hspace{1cm} fluid pressure $\flpr_0(r) \, = \, 
                                            \atmovel^2\, \rho_0(r)/\atmogamma
                                            \, = \, 
                                            \atmovel^2\, \exp(-\atmoalpha\,r)/\atmogamma$} \\ \hline
\end{tabular} 
\end{center}\end{table}

\section{Spline representations from point-wise model $S$}
\label{section:spline-modelS}

We start from the set of point-wise coefficients for the adiabatic 
index $\gamma$ and the density $\rho$, given in the model \texttt{S} 
of \cite{christensen1996current}, from $r=0$ to $r=r_s$. 
Practically, it consists of \num{2482} triples:
\begin{equation}
  (r_k, \, \gamma_k, \, \rho_k)_{k = 1, \ldots, \, \num{2482}}\, ,
   \qquad \text{point-wise representation given in model \texttt{S}} \,.
\end{equation}
From these discrete sets of values, our first task is to 
generate a cubic B-spline model and, because the density 
is exponentially decaying in the atmosphere, we build the 
spline for $\log(\rho)$ instead of $\rho$.
Regarding spline representations, we refer to, e.g., 
\cite[Section 8.3]{NumAKress} and \cite{Piegl2012}.
It defines a representation with piecewise-polynomials 
of order $3$ and we rely on the Matlab routine \texttt{spline} 
which, given a set of positions and associated coefficients, 
generate the spline representation.
Nonetheless, to avoid an oscillatory behaviour, we must not 
use all of the positions given in the point-wise representation.

To evaluate the accuracy of our spline representation compared 
to a given point-wise model $m_k \, = \, \{\gamma_k, \,\,\log(\rho_k)\}$, we 
define the maximal relative error $\epsilon_\infty$ such that 
\begin{equation} \label{eq:model-error-spline}
  \epsilon_\infty(m) \, = \, \max_k \, \, \dfrac{\big\vert m_k \, - \, \mathcal{S}_m^{N_s}(r_k) \big\vert}{\vert m_k \vert} \, ,
\end{equation}
where $\mathcal{S}_m^{N_s}(r_k)$ is the spline representation associated
to $m$ using $N_s$ nodes of the model \texttt{S} and evaluated in $r_k$.
To generate the spline representation, we want to keep $N_s$ as small as
possible, while we impose $\epsilon_\infty < \num{5e-4}$, that is, less
than $\num{0.05}\%$ difference between the spline representation and the 
original point-wise models.
The following procedure is employed for this purpose:
\begin{enumerate} \setlength{\itemsep}{-1pt}
  \item Generate an initial spline representation using \num{37} 
        points (one every \num{70}) of the model \texttt{S}.
  \item Evaluate the resulting spline at all of the positions 
        given in the model \texttt{S}.
  \item Compute the maximal relative error $\epsilon_\infty$ in \cref{eq:model-error-spline}, 
        and identify the interval in which it is contained.
  \item Add \num{8} points in the interval in which the maximal error is 
        contained.
  \item Repeat steps 2--4 until $\epsilon_\infty < 0.05$.
\end{enumerate}

%

Eventually, $\log(\rho)$ is represented by 
\num{72} splines ($N_s = \num{73}$) and 
$\gamma$ is represented by 
\num{85} splines ($N_s = \num{86}$). 
Their formulation can further be retrieved 
in the online repository for the models, 
available at \mlink.

\section{Transition and atmosphere region: $\gamma$ and $\rho_0$}
\label{rep_gamma::sec}

We now construct the functions $\gamma$ and $\rho_0$ in the transition
region, see \cref{figure:illustration}.

\subsection{Representation of $\gamma$ on $(r_s,\infty)$}
\label{subsection:gamma_rs-infty}

We start with the $\mathcal{C}^2$ representation for 
$\gamma $ on $[0, \rs]$, given by the spline model 
built in \cref{section:spline-modelS}. 
Following our atmospheric model \texttt{AtmoI} (see~\cref{table:properties}),
we impose that $\gamma = \atmogamma$ for $r\geq \ra$.
In the transition, we choose the ansatz so that 
$\gamma $ is $\mathcal{C}^2$ globally:
\begin{equation}
  \gamma(r) \, = \, \begin{cases} 
         \text{model \texttt{S}-based splines} \,\, \gammaspline(r) \, &, \, r\in [0, \rs]\,;\\
  f(r)      \,\,\,  &, \,  r \in (\rs, \ra)\,;\\
  \atmogamma \, &, \, r \geq \ra
\end{cases}
\end{equation}
with $f$ of the form, 
\begin{equation}\label{ansatzgamma}
  f(r) \, = \, \atmogamma \, + \, a (r - \ra)^3 \, + \, b (r - \ra)^4\,.
\end{equation}
This ansatz ensures that the function is $\mathcal{C}^2$ 
at $r=\ra$. Next, we impose, at $r = \rs$, 
\begin{equation}
  f (\rs)  \,=\, \gammaspline (\rs) \, , \qquad 
  f'(\rs)  \,=\, \gammaspline'(\rs) \, , \qquad 
  f''(\rs) \,=\, \gammaspline''(\rs)\,.
\end{equation}
Substituting these condition in the ansatz \cref{ansatzgamma}, 
we have the following three equations to identify the three 
unknowns: $a$, $b$ and $\atmogamma$:
\begin{subequations} \label{gamma:system-a-b-infty} \begin{empheq}[left={\empheqlbrace}]{align}
\atmogamma \, + \, a (\rs - \ra)^3 \, + \, b (\rs - \ra)^4 \, &= \, \gammaspline(\rs)\,,\\
                3 a (\rs - \ra)^2 \, + \, 4 b (\rs - \ra)^3 \,& = \, \gammaspline'(\rs)\,,\\
 6 a  ( \rs - \ra) \, + \, 12 b (\rs - \ra)^2  \, &= \, \gammaspline''(\rs)\,.
\end{empheq} \end{subequations}

From the last two equations we obtain
\begin{equation}
  (\rs - \ra) \begin{pmatrix} 3 (\rs - \ra) & 4 (\rs - \ra)^2  \\1 & 2 (\rs - \ra) 
  \end{pmatrix} \begin{pmatrix} a \\b \end{pmatrix} 
   = \begin{pmatrix} \gammaspline'(\rs) \\ \tfrac{1}{6}\gammaspline''(\rs)
     \end{pmatrix}\,.
\end{equation}
Thus, we have
\begin{subequations}
\begin{align}
 \begin{pmatrix} a \\b \end{pmatrix} &=  \dfrac{1}{ 2 (\rs - \ra)^2}\begin{pmatrix} 
 2( \rs  - \ra) & -4 (\rs - \ra)^2\\
 -1             & 3 (\rs - \ra)
 \end{pmatrix} \dfrac{1}{ \rs - \ra} \begin{pmatrix} \gammaspline'(\rs) \\ \tfrac{1}{6}\gammaspline''(\rs)
 \end{pmatrix} \\[0.8em]
& = \begin{pmatrix} 
 2( \rs  - \ra)^{-1}  & -2 (\rs - \ra)^{-1}\\[0.5em]
 -2( \rs - \ra)^{-3} & \tfrac{3}{2}(\rs - \ra)^{-2}           
 \end{pmatrix} 
\begin{pmatrix} \gammaspline'(\rs) \\[0.5em] \tfrac{1}{6}\gammaspline''(\rs)
 \end{pmatrix}\,.
\end{align}
\end{subequations}
The first equation of \cref{gamma:system-a-b-infty} 
gives $\atmogamma$ such that
\begin{equation}
  \atmogamma \,  =\,  -\, a (\rs - \ra)^3 \, - \, b (\rs - \ra)^4 \, + \, \gammaspline(\rs)\,.
\end{equation}

Therefore, we obtain the following formulations 
to compute $a$, $b$ and $\atmogamma$:
\begin{subequations}
 \begin{align}
  a &= \frac{3 \gammaspline'(\rs) - \gammaspline''(\rs) (\rs-\ra)}{3(\rs-\ra)^2}  \,;\\
  b &= \frac{- 2 \gammaspline'(\rs) + \gammaspline''(\rs) (\rs-\ra)}{4(\rs-\ra)^3} \,;\\
       \atmogamma &= \gammaspline(\rs) - \frac{ (\rs-\ra)\gammaspline'(\rs)}{2} + \frac{(\rs-\ra)^2 \gammaspline''(\rs)}{12}.
 \end{align}
\end{subequations}

The first and second-order derivatives 
of $\gamma$ are given analytically by
\begin{equation}
  \gamma' \, = \, \begin{cases} 
  \gammaspline'(r) \, &, \, r\in [0, \rs]\,,\\
  f'(r) \, = \, 3 a (r - \ra)^2 \, + \, 4 b (r - \ra)^3    \, &, \,  r \in (\rs, \ra)\,,\\
    0   \, &, \, r \geq \ra\,,
\end{cases}
\end{equation}
and
\begin{equation}
\gamma'' \, = \, \begin{cases} 
\gammaspline''(r)  &, \, r\in [0, \rs]\,,\\
f''(r) = 6 a (r - \ra) \, + \, 12 b (r - \ra)^2  \, &, \,  r \in (\rs, \ra)\,,\\
0 \, &, \, r \geq \ra\,.
\end{cases}
\end{equation}
We remind the scale height functions, given by
\begin{equation}
  \alpha_\gamma = -\dfrac{\gamma' }{ \gamma}\, , \qquad 
  \alpha'_\gamma = -\dfrac{\gamma'' \gamma - {\gamma'}^2}{ \gamma^2}\,.
\end{equation} 


\subsection{Representation of $\rho_0$ on $(r_s,\infty)$}
\label{rep_rho::sec}

We start with the $\mathcal{C}^2$ representation for 
$\log(\rho_0)$ on $[0, \rs]$, given by the spline model 
built in \cref{section:spline-modelS}, and that we refer
to as $\logrhospline(r)$. 
The atmospheric model \texttt{AtmoI} we have introduced, 
see~\cref{table:properties}, imposes that 
\begin{equation}
 \text{for} \,\,\, r\geq \ra \, , \quad 
 \log \rho_0(r)  = - \atmoalpha (r-\ra) + \log \rho_0(\ra) \, ,
\end{equation}
with $\atmoalpha$ and $\log\rho_0(\ra)$ to be determined.

Look for an ansatz in the transition so that 
$\log(\rho_0)$ is $\mathcal{C}^2$ globally:
\begin{equation}
  \log\rho_0 \, = \, \begin{cases} 
  \, \text{model \texttt{S}-based splines} \,\, \logrhospline(r) &, \, r\in [0, \rs]\,;\\
  f(r)   \, &, \,  r \in (\rs, \ra)\,;\\
  - \atmoalpha (r-\ra) + \log \rho_0(\ra) \, &, \, r \geq \ra \, , 
\end{cases}
\end{equation}
with $f$ of the form, 
\begin{equation}\label{ansatzlogrho}
  f(r) \, = \, \alpha_3 (r - \rs)^3 \, + \, \alpha_2 (r - \rs)^2 \, + \, \alpha_1 (r-\rs) \, + \, \alpha_0.
\end{equation}
As we impose the function $\log(\rho_0)$ to be $\mathcal{C}^2$, we have at $r = \rs$, 
\begin{equation}
  f (\rs)  \,=\, \logrhospline(\rs)   \, , \qquad 
  f'(\rs)  \,=\, \logrhospline'(\rs)  \, , \qquad 
  f''(\rs) \,=\, \logrhospline''(\rs) \, . 
\end{equation}
Substituting these conditions in the ansatz, 
we obtain at $r=\rs$
\begin{subequations}\begin{empheq}[left={\empheqlbrace}]{align}
  f(\rs)   &= \logrhospline  (\rs) \, = \, \alpha_0\,; \\
  f'(\rs)  &= \logrhospline' (\rs) \, = \, \alpha_1 \,;\\
  f''(\rs) &= \logrhospline''(\rs) \, = \, 2 \alpha_2.
\end{empheq}
\end{subequations}

In $r = \ra$, model \texttt{AtmoI} imposes $f''(\ra) = 0$, such that,
\begin{equation}
  f''(\ra) \, = \, 0 \, = \, 6 \alpha_3 (\ra - \rs) 
            + 2 \alpha_2 \quad \Rightarrow \qquad
            \alpha_3 \, = \, - \frac{\alpha_2}{3(\ra-\rs)} \,.
\end{equation}
We also have, in $r=\ra$,
\begin{subequations}
  \begin{align}
   f'(\ra) &\,=\, -\atmoalpha \,=\, 
              3 \, \alpha_3 (\ra-\rs)^2 + 2 \, \alpha_2 (\ra-\rs) + \alpha_1  \,, \\
   f (\ra) &\,=\, \log \rho_0(\ra) = \alpha_3 (\ra - \rs)^3 \, + \, \alpha_2 (\ra - \rs)^2 \, + \, \alpha_1 (\ra-\rs) \, + \, \alpha_0.
  \end{align}
\end{subequations}


The inverse density scale heights function 
$\alpha_{\rho_0} = -(\log \rho_0)'$ is then
obtained analytically:
\begin{equation}
  \alpha_{\rho_0} \, = \, \begin{cases} 
   - \logrhospline'(r)  \, &, \, r\in [0, \rs]\,;\\
   f'(r) \, = \, 3 \, \alpha_3 (r - \rs)^2 \, + \, 2 \alpha_2 (r - \rs) 
              \, + \, \alpha_1   \, &, \,  r \in (\rs, \ra)\,;\\
   - \atmoalpha \, &, \, r \geq \ra \,,
\end{cases}
\end{equation}
and 
\begin{equation}
  \alpha_{\rho_0} \, = \, \begin{cases} 
   - \logrhospline''(r)  \, &, \, r\in [0, \rs]\,;\\
   f''(r) \, = \, 6 \, \alpha_3 (r - \rs) \, + \, 2 \alpha_2
               \, &, \,  r \in (\rs, \ra)\,;\\
   0 \, &, \, r \geq \ra \,.
\end{cases}
\end{equation}

\section{Pressure, gravitational potential and velocity}
\label{rep_p::sec}

The model \texttt{S} prescribes point-wise values of the 
fluid pressure $\flpr_0$, similarly as for the density and the 
adiabatic index. However, to ensure that the hydrostatic equilibrium 
\cref{hydrostatic} is strictly preserved, the pressure is 
instead retrieved from the representations of $\rho_0$ and $\gamma$ 
we have introduced above.
Hence we shall only use from the model \texttt{S} the value of 
$\flpr_0$ in $\rs$, see \cref{subsection:pressure}.

\subsection{Computation of the derivatives of the gravitational potential}
\label{subsection:compute-gravity-pot}

From $\rho_0(r)$ computed above, the first and second-order 
derivatives of the gravitational potential, $\Phi_0'$ and $\Phi_0''$
are given by,
\begin{subequations}\begin{align}
  \Phi_0'(r)  &\,=\, \frac{4\pi G}{r^2} \int_0^r \rho_0(s) s^2 \mathrm{d}s\,, \label{eq:dphi0_integral}\\
  \Phi_0''(r) &\,=\, 4\pi \,G\, \rho_0(r) \,-\, \frac{2}{r} \Phi_0'(r)\,.
\end{align}\end{subequations}
The functions $\Phi_0'$ and $\Phi_0''$ 
are continuous at $r=0$ with
\begin{subequations}\begin{align}
  \Phi_0'(0)  &\,=\, \lim_{r\rightarrow 0} \,\Phi_0'(r) \,=\, \lim_{r\rightarrow 0} \frac{4\pi G}{3} r \rho \,=\, 0\,; \\
  \Phi_0''(0) &\,=\, 4\pi \,G\, \rho(0) \,-\,\frac{8\pi G \rho(0)}{3} \,=\, \frac{4\pi G \rho(0)}{3}\,.
\end{align}\end{subequations}

\begin{rmk}
  We note that the third-order derivative is given by,
  \begin{equation}
     \Phi_0'''(r) \,=\, 4\pi G \rho_0'(r) \,-\, 2 \frac{\Phi_0''(r)}{r} \,+\, 2 \frac{\Phi_0'(r)}{r^2} \,,
  \end{equation}
  and using the previous limits
  \begin{equation}
  \Phi_0'''(0) \,=\, 4\pi G \rho_0'(0)\,.
  \end{equation}
  Consequently, $\Phi_0$ is at least $\mathcal{C}^3$ (if $\rho_0$ is at least $\mathcal{C}^1$).
  \hfill $\triangle$
\end{rmk}

\subsection{Computation of the pressure in the interior}
\label{subsection:pressure}

From $\Phi_0'$ and $\rho_0$, the first and second-order derivative 
of the pressure, $\flpr'_0$ and $\flpr_0''$ in the interior are obtained 
using the hydrostatic equilibrium (see \cref{table:properties}): 
\begin{subequations} \label{eq:pressure-deriv}
\begin{align}
  \flpr_0'(r)  & \,=\, - \rho_0(r)\, \Phi'_0(r), \\
  \flpr_0''(r) & \,=\, - \rho_0(r)\, \Phi_0''(r) \,-\, \rho_0'(r) \Phi_0'(r)\,.
\end{align}\end{subequations}
We integrate to obtain $\flpr_0$:
\begin{equation} \label{eq:pressure-int}
 \flpr_0(r) = \flpr_0(\rs) + \int_r^{\rs} \rho_0(s) \Phi'_0(s) \mathrm{d}s.
\end{equation}
The value of $\flpr_0(\rs)$ is given by the data point from 
model \texttt{S} of \cite{christensen1996current}.

\subsection{Computation of the pressure in the transition and atmosphere}

For the computation of the pressure in the transition 
region and the atmosphere, we take
\begin{subequations}
\begin{align}
\flpr_0'(r) \, &= \, - \rho_0(r) \Phi_0'(r)\, 
+ \,  \tEhe(r) \, , \qquad r \geq \rs \,; \label{eq:dflpr0:atmo}\\
\flpr_0(r)  \,&= \, \flpr_0(\ra) \, e^{-\atmoalpha(r-\ra)} \, , \hspace*{1.78cm} r\geq \ra\,. \label{eq:flpr0:atmo}
\end{align}
\end{subequations}
That is, in the transition region, the hydrostatic 
equilibrium is no more respected, and $\tEhe$ represents
the distance to this equilibrium.
We write
\begin{subequations}\label{p_0transition}
\begin{align}
  \flpr_0(r) &= - \mathfrak{M}(r) + Q(r) + \flpr_0(\rs) \,, \qquad\qquad\qquad\qquad r\in (\rs, \ra)\,,\\[0.5em]
  \flpr_0(r) &= e^{-\atmoalpha( r - \ra)}\,\, \lim_{r\rightarrow \ra^-}\left( - \mathfrak{M}(r) + Q(r) 
              + \flpr_0(\rs)     \right)    \,\,    
              \,, \qquad r\geq \ra \, ,
\end{align}
\end{subequations}
with
\begin{equation}\label{MQdef}
\mathfrak{M}(r):= \int_{\rs}^r \rho_0 \, \Phi_0' \, ds
\, , \qquad Q(r) := \int_{\rs}^r \tEhe\,.
\end{equation}

For the function $\tEhe$, we consider on $[\rs, \ra]$ 
that it is a polynomial, with the ansatz, 
\begin{equation}\label{ansatztEhe}
  \tEhe \, =\,  ( r - \rs)^3   \sum_{n=0}^N   a_n ( r - \ra)^n\,.
\end{equation}
Here, $\tEhe = 0$ in $r=\rs$, where the hydrostatic
equilibrium still prevails.
The chosen integration factors in \cref{MQdef,p_0transition} 
guarantee the continuity of $\flpr_0$ at $r = \rs$, while
the factor $(r-\rs)^3$ guarantees the continuity 
at $r = \rs$ of $\flpr_0'$, $\flpr_0''$ and $\flpr_0'''$, thus 
$\flpr_0$ is $\mathcal{C}^3$ at $r = \rs$.
It remains to impose the continuity at $r=\ra$ for the
derivatives of $\flpr_0$. We have, using \cref{eq:flpr0:atmo},
\begin{subequations}\label{C3pra}\begin{empheq}[left={\empheqlbrace}]{align}
\flpr_0'(\ra)   \,&= \, -\atmoalpha\,\, \flpr_0(\ra) \,, \label{dpra}\\
\flpr_0''(\ra)  \,&= \,  \phantom{+} \atmoalpha^2 \, \,\flpr_0(\ra)\,,\label{d2pra}\\
\flpr_0'''(\ra) \,&= \,-\atmoalpha^3 \, \,\flpr_0(\ra)\,.\label{d3rpa}
\end{empheq}
\end{subequations}
This amounts to three equations and therefore we have 
to take $N+1=3$ unknowns so $N=2$ in \cref{ansatztEhe}.

The computation follows the steps given below, 
for which we introduce the notation,
\begin{equation}
  \mathfrak{a} := (\rho_0 \Phi_0')(\ra)
  \,, \qquad \mathfrak{b} := (\rho_0 \Phi_0')'(\ra)
  \,, \qquad \mathfrak{c} := (\rho_0 \Phi_0')''(\ra)\,.
\end{equation}

\textbf{Step 1} 
At $r = \ra$, we have
\begin{equation}\begin{aligned}
  \tEhe(\ra) &= (\ra - \rs)^3 a_0 \, ,\\
  \tEhe'(\ra) &= 3 (\ra  - \rs)^2  a_0  + (\ra - \rs)^3 a_1\,, \\
  \tEhe''(\ra)
 &= 6 (\ra - \rs) a_0
  + 6 (\ra - \rs)^2  a_1 + 2 (\ra - \rs)^3 a_2\,.
\end{aligned}
\end{equation}

Using \cref{C3pra}, we obtain
\begin{equation}\label{ratioC3pra}
  \dfrac{\flpr_0'(\ra) }{ -\atmoalpha} \,=\, \dfrac{\flpr_0''(\ra) }{ \,\atmoalpha^2}
  \,, \qquad \dfrac{\flpr_0'(\ra) }{ -\atmoalpha} \,  = \, \dfrac{\flpr_0'''(\ra) }{-\alpha^3_{\infty}}\,.
\end{equation}
The first equation in \cref{ratioC3pra} gives
\begin{equation}
\begin{aligned}
&\dfrac{ - \mathfrak{a}  +(\ra - \rs)^3 a_0 }{ -\atmoalpha}
 \, = \, \dfrac{-\mathfrak{b} + 3 (\ra  - \rs)^2  a_0  + (\ra - \rs)^3 a_1 }{ \,\atmoalpha^2}\,\\[0.5em]
\Rightarrow \qquad a_1  &= \left(\dfrac{ - \mathfrak{a}  +(\ra - \rs)^3 a_0 }{ -\atmoalpha}
 \, -\, \dfrac{-\mathfrak{b} + 3 (\ra  - \rs)^2  a_0   }{ \,\atmoalpha^2} \right)\dfrac{ \atmoalpha^2}{(\ra - \rs)^3 }\,,
 \end{aligned}
 \end{equation}
 and we obtain
 \begin{equation}\label{a1a0}
 a_1\,= \,\dfrac{\mathfrak{a}\, \atmoalpha + \mathfrak{b}}{(\ra - \rs)^3 }
 - a_0 \left(  \atmoalpha  +  \dfrac{3}{( \ra - \rs)}\right) \,.
\end{equation}
The second equation in \cref{ratioC3pra} gives, 
\begin{equation}
\begin{aligned}
& \dfrac{ - \mathfrak{a}  +(\ra - \rs)^3 a_0 }{ \atmoalpha}
 \, = \, \dfrac{- \mathfrak{c}\, + \, 6 (\ra - \rs) a_0
  + 6 (\ra - \rs)^2  a_1 + (\ra - \rs)^3 2a_2 }{ \atmoalpha^3}\\
  \Rightarrow\qquad &
\dfrac{-\mathfrak{a}\, \atmoalpha^2+ \mathfrak{c}}{2(\ra - \rs)^3}
\, + \, \left(\dfrac{\atmoalpha^2}{2} - \dfrac{ 3}{(\ra - \rs)^2}\right) a_0
 - \dfrac{ 3}{(\ra - \rs)} a_1 
\, = \,  a_2\,.
 \end{aligned}
\end{equation}
Substitute $a_1$ in terms of $a_0$ using \cref{a1a0}, we get
\begin{equation}
\begin{aligned}
a_2 &= \dfrac{-\mathfrak{a}\, \atmoalpha^2+ \mathfrak{c}}{2(\ra - \rs)^3}
\, + \, \left(\dfrac{\atmoalpha^2}{2} - \dfrac{ 3}{(\ra - \rs)^2}\right) a_0
 - \dfrac{ 3}{(\ra - \rs)} \left( \dfrac{\mathfrak{a}\, \atmoalpha + \mathfrak{b}}{(\ra - \rs)^3 }
 - a_0 \left(  \atmoalpha  +  \dfrac{3}{( \ra - \rs)}\right)\right)\\
 &= \left(\dfrac{-\mathfrak{a}\, \atmoalpha^2+ \mathfrak{c}}{2(\ra - \rs)^3}- \dfrac{ 3}{(\ra - \rs)}   \dfrac{\mathfrak{a}\, \atmoalpha + \mathfrak{b}}{(\ra - \rs)^3 } \right)\\
&\hspace*{2cm} \, + \, a_0 \left(\dfrac{\atmoalpha^2}{2} - \dfrac{ 3}{(\ra - \rs)^2} \, + \, \dfrac{ 3}{(\ra - \rs)}  \left(  \atmoalpha  + \dfrac{3}{( \ra - \rs)}\right)    \right)\,,
\end{aligned}
\end{equation}
and
\begin{equation}\label{a2a0}
\begin{aligned}
a_2 \, =\, & \left(-\mathfrak{a} \left(\dfrac{\alpha^2_{\infty}}{2}+ \dfrac{ 3 \atmoalpha }{(\ra - \rs)}\right)  
+ \dfrac{\mathfrak{c}}{2}
- \dfrac{ 3 \mathfrak{b}}{(\ra - \rs)}    \right)\dfrac{1}{(\ra - \rs)^3} \\
& \, + \, 
  a_0 \left(\dfrac{\atmoalpha^2}{2} + \dfrac{ 6}{(\ra - \rs)^2} \, + \, \dfrac{ 3 \atmoalpha}{(\ra - \rs)}   \right)\,.
\end{aligned}
\end{equation}
For a compact notation, we define the constants 
in \cref{a1a0,a2a0} using $C_{ij}$, giving
\begin{equation}\label{Cijdef}
a_1 \, = \, C_{10}\,\, +\,\, C_{11}\,\, a_0 \, , \qquad
a_2 \, = \, C_{20} \,\, + \,\,C_{21} \,\,a_0\,.
\end{equation}

\textbf{Step 2}
We rewrite $\tEhe$ as, 
\begin{equation}
\begin{aligned}
\tEhe 
&= ( r - \rs)^3   \left(   a_0 + a_1 ( r -  \ra) + a_2( r - \ra)^2\right)\\
&= ( r - \rs)^3   \big(   a_0 + a_1 ( r -  \rs) + a_1( \rs - \ra)\, + \, a_2( (r -\rs)^2 \\ 
                  & \hspace*{4.60cm}  + 2 (r-\rs)( \rs - \ra) \, + \, (\rs - \ra)^2\big)\\
&=( r - \rs)^3   \big(   a_0 + a_1( \rs - \ra)  +  a_2 (\rs - \ra)^2 + a_1 ( r -  \rs) \, \\
                 & \hspace*{4.75cm}   + a_2 (r -\rs)^2 + 2 a_2 ( \rs - \ra) (r-\rs)\big)\\
&= ( r - \rs)^3 \left( C + B ( r - \rs)  +  a_2 (r -\rs)^2 \right)\, ,
\end{aligned}
\end{equation}
where
\begin{equation}\label{tempnotationCB}
C\,  :=\,   a_0 + a_1( \rs - \ra)  +  a_2 (\rs - \ra)^2
\, , \qquad B:= a_1 +  2 a_2 ( \rs - \ra)\,.
\end{equation}
From the definition of $Q$ in \cref{MQdef}, we have
\begin{equation}
\begin{aligned}
Q(r) &:= \int_{\rs}^r \tEhe\, \mathrm{d}s
 = \int_{\rs}^r    C( r - \rs)^3 + B ( r - \rs)^4  +  a_2 (r -\rs)^5 \,\, \mathrm{d} s\\
 & = \dfrac{C}{4}( r - \rs)^4 + \dfrac{B}{5} ( r - \rs)^5  + \dfrac{ a_2}{6} (r -\rs)^6 \,.
\end{aligned}
\end{equation}
Substitute the definition of $B$ and $C$ into $Q$,
\begin{equation}
\begin{aligned}
Q(\ra) 
 &= \dfrac{a_0 + a_1( \rs - \ra)  +  a_2 (\rs - \ra)^2}{4}( \ra - \rs)^4 \\
    & \hspace*{3cm} + \dfrac{a_1 +  2 a_2 ( \rs - \ra)}{5} ( \ra - \rs)^5  + \dfrac{ a_2}{6} (\ra -\rs)^6\\
&= \dfrac{( \ra - \rs)^4}{4} a_0
\, + \, a_1 ( \ra - \rs)^5 \left( -\dfrac{1}{4}+ \dfrac{1}{5}\right)
\,+ \, a_2 ( \ra - \rs)^6 \left(\dfrac{1}{4} - \dfrac{2}{5}  + \dfrac{1}{6}\right)\\
&= \dfrac{( \ra - \rs)^4}{4} a_0
\, - \, a_1 ( \ra - \rs)^5 \dfrac{1}{20}
\,+ \, a_2 ( \ra - \rs)^6 \dfrac{1}{60} \,.
\end{aligned}
\end{equation}
For simplicity, this last expression is written as
\begin{equation}
Q(\ra) \,=\,  Q_0 \,\,a_0\,\, +\,\, Q_1  \,\,a_1 \,\,+\,\, Q_2 \,\,a_2\,.
\end{equation}
In terms of $C_{ij}$ defined in \cref{Cijdef}, we have
\begin{equation}
\begin{aligned}
Q(\ra) \,\,
 = \,\,Q_0 a_0 + Q_1 (C_{10} + C_{11} a_0) 
 + Q_2 (C_{20}  + C_{21} a_0 )\\
\,\,  =\,\, (Q_0 + Q_1 C_{11} + Q_2 C_{21} ) a_0 +Q_1 C_{10} + Q_2 C_{20}\,.
\end{aligned}
  \end{equation}

We now return to equation \cref{dpra}, we have
$ \flpr_0'(\ra) = -\atmoalpha \flpr_0(\ra)$, 
\begin{equation}
\dfrac{ - \mathfrak{a}  +(\ra - \rs)^3 a_0 }{ -\atmoalpha}
 = - \mathfrak{M}(\ra) \, + \, (Q_0 + Q_1 C_{11} + Q_2 C_{21} ) a_0 +Q_1 C_{10} + Q_2 C_{20}
  \, + \, \flpr_0(\rs)\,,
\end{equation}
and solve for $a_0$,
\begin{equation} \label{eq:a0}
a_0 \left( Q_0 + Q_1 C_{11} + Q_2 C_{21} + \dfrac{ (\ra - \rs)^3}{ \atmoalpha}\right)  =  \mathfrak{M}(\ra) + \dfrac{\mathfrak{a}}{\atmoalpha}  -Q _1 C_{10} - Q_2 C_{20}
  \, - \, \flpr_0(\rs)\,.
\end{equation}

%
%

\medskip

In the transition region, we compute $\flpr'_0$ using \cref{eq:dflpr0:atmo}, where 
\begin{equation}\label{tEhe}
  \tEhe \, =\,  ( r - \rs)^3   \left(   a_0 + a_1 ( r - \ra) + a_2 (r-\ra)^2 \right)\,,
  \qquad r \in (\rs, \ra) \, ,
\end{equation}
and the constant $a_0$ is given by \cref{eq:a0} and $a_1$ and $a_2$ are obtained from $a_0$ using \cref{a1a0,a2a0}. We can obtain the next derivatives of the pressure
\begin{subequations} 
\begin{empheq}[left={r \in (\rs, \ra) \qquad\empheqlbrace}]{align}
\flpr_0''(r) \,&= \, -\rho_0'(r) \Phi_0'(r) - \rho_0(r) \Phi_0''(r) + \tEhe' \, , \label{eq:ddflpr}\\
\flpr_0'''(r) \,&= \, -\rho_0''(r) \Phi_0'(r) - 2 \rho'_0(r) \Phi_0''(r) -\rho_0(r) \Phi_0'''(r) +  \tEhe'' \label{eq:dddflpr}\,,
\end{empheq}
\end{subequations}
where the derivatives of $\tEhe$ are obtained analytically from \cref{tEhe}.

In the atmosphere, $\flpr_0$ is given by \cref{eq:flpr0:atmo} and the derivatives are thus
\begin{subequations} \label{eq:deriv_flpr_atmo}
\begin{empheq}[left={r \geq \ra \qquad\empheqlbrace}]{align}
\flpr_0'(r) \,&= \, - \atmoalpha \flpr_0(\ra) \, e^{-\atmoalpha(r-\ra)} \, ,\\
\flpr_0''(r) \,&= \, \atmoalpha^2  \flpr_0(\ra) \, e^{-\atmoalpha(r-\ra)} \, , \\
\flpr_0'''(r) \,&= \, -\atmoalpha^3  \flpr_0(\ra) \, e^{-\atmoalpha(r-\ra)} \,.
\end{empheq}
\end{subequations}
Then, the expression of $\tEhe$ is obtained from \cref{eq:dflpr0:atmo} and its 
derivatives using \cref{eq:ddflpr,eq:dddflpr}.

\subsection{Velocity and inverse scale height functions}

The velocity is obtained from the adiabaticity (\cref{table:properties})
from $\rho_0$, $\gamma$ and $\flpr_0$, such that,
\begin{equation}
 \vel \, = \,  \sqrt{\frac{\gamma \,\, \flpr_0}{\rho_0}}.
\end{equation}

Furthermore, the inverse scale height functions 
$\alpha_\bullet$ are given by
\begin{subequations}\begin{align}
 \alpha_{\gamma \flpr_0}   &\,=\, \alpha_\gamma \,+\, \alpha_{\flpr_0}, \\
 \alpha_{\vel}             &\,=\, \frac{1}{2} \left( \alpha_{\flpr_0} \,-\, \alpha_{\rho_0} \,+\, \alpha_\gamma \right), \\
 \alpha'_{\gamma \flpr_0}  &\,=\, \alpha'_\gamma \,+\, \alpha_{\flpr_0} \,+\, \alpha_\gamma \,+\, \alpha'_{\flpr_0}, \\
 \alpha'_{\vel}            &\,=\, \frac{1}{2} \left( \alpha'_{\flpr_0} \,-\, \alpha'_{\rho_0} \,+\, \alpha'_\gamma \right).
\end{align}\end{subequations}


\section{Summary of important values}

In this section, we review the methodology to generate the 
solar models, and explicitly give some key-values we obtain
for the models that are made available at \mlink.
From the computational steps we have prescribed, the background 
models depend \emph{only} on the following choices: 
\begin{enumerate}
  \item The choice of $\rs$: we use the last entry given in  
        the model \texttt{S}: 
       \begin{equation}
       \rs  \,=\, \num{1.000716} \,. 
       \end{equation}
  \item The choice of $\ra$: we consider that the atmosphere starts in
       \begin{equation}
       \ra  \, = \, \num{1.00073} \,. 
       \end{equation}
  \item The selection of nodes for the spline representation of $\log(\rho_0)$ 
        and $\gamma$, for which we follow the procedure given in \cref{section:spline-modelS}.
  \item The choice of approximation for the numerical integration to compute 
        \cref{eq:dphi0_integral,eq:pressure-int}: we use a trapeze rule
        with a discretization step $\num{e-7}$.
\end{enumerate}
We review the computational steps in \cref{algo:main}, while
the resulting background solar models and scripts to generate
them are available at \mlink.
We note that 
\begin{equation}
  \dfrac{r}{\vel(r)} \qquad \text{ is decreasing on }  [0,\infty) \,,
\end{equation}
\begin{equation}
  \text{for $r \geq \ra$}, \qquad
  \,\alpha_{\rho_0} = \alpha_{\flpr_0}  = \alpha_{\gamma \flpr_0} = \atmoalpha\,, 
  \qquad\qquad \vel  = \atmovel \,,\qquad\qquad \gamma = \atmogamma\,.
\end{equation}
We review in \cref{table:gamma,table:rho,table:vel,table:ehe,table:pressure} 
the main background parameters.

\newlength{\lstep}
\setlength{\lstep}{-0.4em}
\begin{algorithm}[ht!]
 \renewcommand{\arraystretch}{1.2}
 \begin{mdframed}
 \KwData{point-wise models for the density and adiabatic index.}
 \KwData{choice of $\rs$ and $\ra$.}
 \vspace*{-\lstep}

  1. Compute piecewise-polynomial representation for $\gamma$ and $\log(\rho_0)$:
  \vspace*{-.3em}

  \begin{itemize} \setlength{\itemsep}{-2pt}
    \item spline representation in $[0, \, \rs]$ from points given in 
          model \texttt{S}, cf. \cref{section:spline-modelS},
    \item extension from $[\rs, \, \infty]$, cf \cref{rep_gamma::sec},
    \item the scale height functions and derivatives follow explicitly.
  \end{itemize}
  \vspace*{-\lstep}
  
  2. Compute the derivatives of the background gravity potential   
     $\Phi_0'$, $\Phi_0''$ and $\Phi_0'''$  in $[0, \, \infty]$, cf.~\cref{subsection:compute-gravity-pot}. \\[\lstep]


  3. Compute the pressure and its derivatives $\flpr_0$, $\flpr_0'$,  $\flpr_0''$
     and $\flpr_0'''$ in $[0, \, \rs]$, cf.~\cref{subsection:pressure}. \\[\lstep]

  4. Compute $\tEhe$, in $[\rs, \, \ra]$ (it is $0$ in $[0, \, \rs]$), 
     it is a polynomial of order $\num{5}$, cf.~\cref{tEhe}. \\[\lstep]

  5. From $\tEhe$, we obtain $\flpr_0'$ in $[\rs,\ra]$ using  \cref{p_0transition} and the derivatives $\flpr_0''$ and $\flpr_0'''$ using \cref{eq:ddflpr,eq:dddflpr} as well as $\flpr_0$ by integration. \\[\lstep]

  6. Compute the pressure $\flpr_0$ in $[\ra, \, \infty]$ using ~\cref{eq:flpr0:atmo} and its derivatives using \cref{eq:deriv_flpr_atmo}. \\[\lstep]

  7. Compute $\tEhe$ in $[\ra, \, \infty]$ using \cref{eq:dflpr0:atmo} and its derivatives from \cref{eq:ddflpr,eq:dddflpr}. \\[\lstep]

  8. Compute $\Ehe$ and the auxiliary functions ($\vel$, scale heights) and derivatives. \\[\lstep]
 \end{mdframed}
 
 \caption{Steps for the computation of $\mathcal{C}^2$ solar models.}
 \label{algo:main}
\end{algorithm}

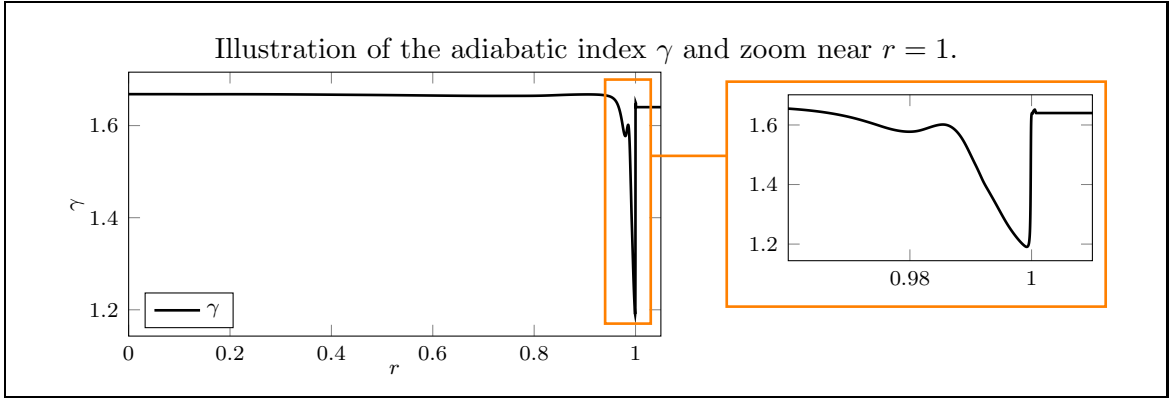
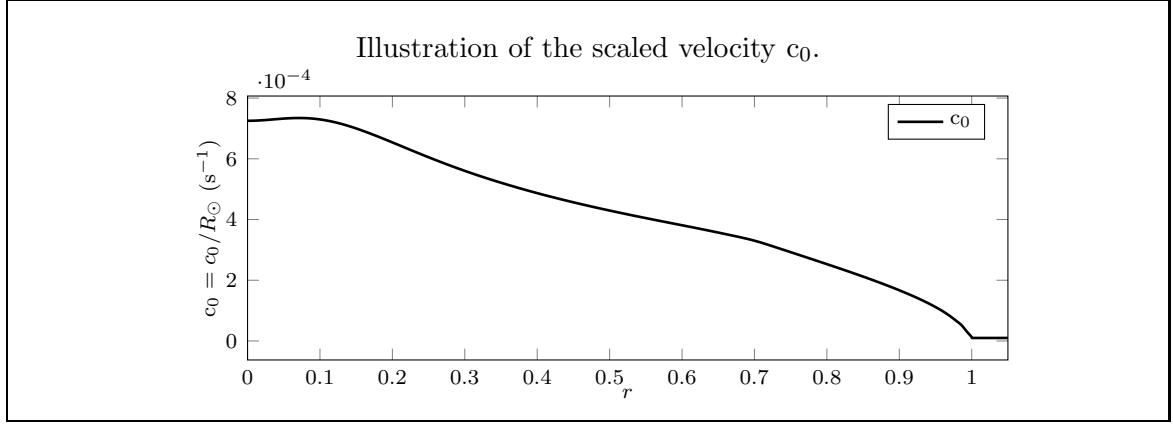
\setlength{\plotwidth} {10.00cm}
\setlength{\plotheight} {4.00cm}

\begin{table}[ht!] \begin{center}
\caption{Summary of information for the adiabatic index $\gamma$.}
\label{table:gamma}
\renewcommand{\arraystretch}{1.20}
\begin{tabular}{|>{\raggedright\arraybackslash}p{.08\linewidth}|
                 >{\raggedright\arraybackslash}p{.85\linewidth}|} 
\hline 
  \multicolumn{2}{|c|}{\textbf{adiabatic index} $\gamma$} \\ \hline
  $[0,\, \rs]$   & principal parameter defined using a spline representation 
                   from the point-wise values given in model \texttt{S}. In $\rs$,
                   we have, \\ 
                 & $\gamma(\rs)=$ \num{1.6407053}.     \\[.5em] \hdashline
  $[\rs, \, \ra]$& $\gamma(r) = \atmogamma + a(r - \ra)^3 + b(r - \ra)^4 \, ,$ \\
                 & with $\qquad a=$ \num{-2.3822371e11}, $\qquad b=$ \num{-6.8240786e15}. \\[.5em] \hdashline
  $r > \ra$      & constant value $\gamma(r) \,=\, \atmogamma=$ \num{1.6400759} \\ 
  \hline \hline
  \multicolumn{2}{|c|}{} \\[-0.6em]
  \multicolumn{2}{|c|}{Illustration of the adiabatic index $\gamma$ 
                       and zoom near $r=1$.} \\
  \multicolumn{2}{|l|}{
  \renewcommand{\datafile} {models_main.txt}
  \renewcommand{\dataA}    {gamma} 
  \renewcommand{\legendA}  {$\gamma$} 
  \renewcommand{\myylabel} {$\gamma$} 
  \renewcommand{\legendpos}{south west}
  \renewcommand{\zxmin}    {0.96}   \renewcommand{\zxminbox} {0.94} 
  \renewcommand{\zxmax}    {1.01}   \renewcommand{\zxmaxbox} {1.03} 
  \renewcommand{\zymin}    {1.17}   \renewcommand{\zymax}    {1.70}   
  \renewcommand{\xticka}   {0.98}   \renewcommand{\xtickb}   {1}
  \hspace*{-4em}
  \setlength{\plotwidth}  {7.00cm} \setlength{\plotheight} {3.50cm}
\begin{tikzpicture}
\begin{axis}[width=\plotwidth, height=\plotheight,
               scale only axis, yminorticks=true,clip mode=individual,
               enlarge y limits=true,
               enlarge x limits=false,
               y label style={xshift=0cm, yshift=-0.5em},
               ylabel={\myylabel},xlabel={$r$},
               x label style={xshift=0cm, yshift=0.20cm},
               legend pos=\legendpos,
               label style     ={font=\scriptsize},
               tick label style={font=\scriptsize},
               legend style    ={font=\scriptsize\selectfont},
               legend cell align={left}
               ]
     \pgfmathsetmacro{\scale}{1} 
     \addplot[color=black,line width=1]  
              table[x=x,y expr=\scale*\thisrow{\dataA}]
              {\datafile}; \addlegendentry{\legendA} 

     \draw[line width = 1,orange] (\zxminbox,\zymin) 
     -- (\zxmaxbox,\zymin)-- (\zxmaxbox,\zymax)-- (\zxminbox,\zymax) -- cycle ;

\end{axis}
\end{tikzpicture}
  \hspace*{-0.55cm} 
 {\raisebox{3cm}{\begin{tikzpicture}[]
    \coordinate (d1)   at (0,0);
    \draw[line width = 1,orange] ([xshift=0mm,yshift=0mm] d1) -- ([xshift=-1cm,yshift=0mm] d1);
  \end{tikzpicture}}}
  \setlength{\plotwidth}  {4.00cm}  \setlength{\plotheight}  {2.20cm}
  \renewcommand{\datafile} {models_main_zoom.txt}
  \hspace*{-5.5mm}{\raisebox{1cm}{
\begin{tikzpicture}[framed,background rectangle/.style={line width=1,draw=orange}]]
\begin{axis}[width=\plotwidth, height=\plotheight,
               scale only axis, yminorticks=true,clip mode=individual,
               enlarge y limits=true,
               enlarge x limits=false,
               xmin=\zxmin,xmax=\zxmax,
               xtick={\xticka, \xtickb},
               xticklabels={\num{\xticka},\num{\xtickb}},
               xlabel={},
               x label style={xshift=0cm, yshift=0.20cm},
               label style     ={font=\scriptsize},
               tick label style={font=\scriptsize},
               legend style    ={font=\scriptsize\selectfont},
               legend cell align={left}
               ]
     \pgfmathsetmacro{\scale}{1} 
     \addplot[color=black,line width=1]  
              table[x=x,y expr=\scale*\thisrow{\dataA}]
              {\datafile}; 
\end{axis}
\end{tikzpicture}}}
  } \\ \hline
\end{tabular}
\end{center} \end{table}

\begin{table}[ht!] \begin{center}
\caption{Summary of information for the density.}
\label{table:rho}
\renewcommand{\arraystretch}{1.20}
\begin{tabular}{|>{\raggedright\arraybackslash}p{.08\linewidth}|
                 >{\raggedright\arraybackslash}p{.85\linewidth}|}
\hline
\multicolumn{2}{|c|}{\textbf{density $\rho_0$ and it inverse scale height $\alpha_\rho$}} \\ \hline
  $[0,\, \rs]$   & principal parameter defined using a spline representation 
                   from the point-wise values given in model \texttt{S}. In $\rs$,
                   we have, \\ 
                 & $\rho_0(\rs)=$ \num{3.2924832e-9} \si{\g\per\cm\cubed},   
                 \\[.5em] \hdashline
  $[\rs, \, \ra]$& $\log \rho_0(r)  =   \alpha_3 (r - \rs)^3 \, + \, \alpha_2 (r - \rs)^2 
                    \, + \, \alpha_1 (r-\rs) \, + \, \alpha_0 \, ,$ \\
                 & with $\qquad \alpha_0 = \num{-1.9531623e1}$ \si{\g\per\cm\cubed},
                   $\quad \alpha_1 = \num{-6.6335853e3}$ \si{\g\per\cm\cubed},\\
                 & $\hspace*{1.68cm} \alpha_2 = \phantom{+}\num{6.4022322e4}$  \si{\g\per\cm\cubed},
                   $\quad \alpha_3 = \num{-1.2264813e9}$ \si{\g\per\cm\cubed}. \\[.5em] \hdashline
  $r > \ra$      & $\rho_0(r) \,=\, \exp(- \atmoalpha \, r) \, ,$\\
                 & with $\atmoalpha=$ \num{6.6324713e3}.         \\ 
  \hline \hline
  \multicolumn{2}{|c|}{} \\[-0.6em]
  \multicolumn{2}{|c|}{Illustration of the inverse density scale height $\alpha_\rho$ 
                       on a logarithmic scale.} \\
  \multicolumn{2}{|l|}{
  \renewcommand{\datafile} {models_main.txt}
  \renewcommand{\dataA}    {alpha} 
  \renewcommand{\legendA}  {$\alpha_\rho$} 
  \renewcommand{\myylabel} {$\alpha_\rho$} 
  \renewcommand{\legendpos}{south west}
  \renewcommand{\zxmin}    {0.9975}  \renewcommand{\zxminbox} {0.94} 
  \renewcommand{\zxmax}    {1.0015}  \renewcommand{\zxmaxbox} {1.02} 
  \renewcommand{\zymin}    {10}      \renewcommand{\zymax}    {10000}   
  \renewcommand{\xticka}   {0.998}   \renewcommand{\xtickb}   {1.001}
  \hspace*{-4em}
  \setlength{\plotwidth}  {7.00cm} \setlength{\plotheight} {3.50cm}
\begin{tikzpicture}
\begin{axis}[width=\plotwidth, height=\plotheight,
               scale only axis, yminorticks=true,clip mode=individual,
               enlarge y limits=true,
               enlarge x limits=false,
               ymode=log,
               y label style={xshift=0cm, yshift=-0.5em},
               ylabel={\myylabel},xlabel={$r$},
               x label style={xshift=0cm, yshift=0.20cm},
               legend pos=\legendpos,
               label style     ={font=\scriptsize},
               tick label style={font=\scriptsize},
               legend style    ={font=\scriptsize\selectfont},
               legend cell align={left}
               ]
     \pgfmathsetmacro{\scale}{1} 
     \addplot[color=black,line width=1]  
              table[x=x,y expr=\scale*\thisrow{\dataA}]
              {\datafile}; \addlegendentry{\legendA} 

     \draw[line width = 1,orange] (\zxminbox,\zymin) 
     -- (\zxmaxbox,\zymin)-- (\zxmaxbox,\zymax)-- (\zxminbox,\zymax) -- cycle ;

\end{axis}
\end{tikzpicture}
  \hspace*{-0.65cm} 
 {\raisebox{3cm}{\begin{tikzpicture}[]
    \coordinate (d1)   at (0,0);
    \draw[line width = 1,orange] ([xshift=0mm,yshift=0mm] d1) -- ([xshift=-1cm,yshift=0mm] d1);
  \end{tikzpicture}}}
  \setlength{\plotwidth}  {4.00cm}  \setlength{\plotheight}  {2.20cm}
  \renewcommand{\datafile} {models_main_zoom.txt}
  \hspace*{-5.5mm}{\raisebox{1cm}{
\begin{tikzpicture}[framed,background rectangle/.style={line width=1,draw=orange}]]
\begin{axis}[width=\plotwidth, height=\plotheight,
               scale only axis, yminorticks=true,clip mode=individual,
               enlarge y limits=true,
               enlarge x limits=false,
               ymode=log,
               xmin=\zxmin,xmax=\zxmax,
               xtick={\xticka, \xtickb},
               xticklabels={\num{\xticka},\num{\xtickb}},
               xlabel={},
               x label style={xshift=0cm, yshift=0.20cm},
               label style     ={font=\scriptsize},
               tick label style={font=\scriptsize},
               legend style    ={font=\scriptsize\selectfont},
               legend cell align={left}
               ]
     \pgfmathsetmacro{\scale}{1} 
     \addplot[color=black,line width=1]  
              table[x=x,y expr=\scale*\thisrow{\dataA}]
              {\datafile}; 
\end{axis}
\end{tikzpicture}}}
  } \\ \hline
\end{tabular}
\end{center} \end{table}

\begin{table}[ht!] \begin{center}
\caption{Summary of information for the scaled velocity.}
\label{table:vel}
\renewcommand{\arraystretch}{1.20}
\begin{tabular}{|>{\raggedright\arraybackslash}p{.08\linewidth}|
                 >{\raggedright\arraybackslash}p{.85\linewidth}|}
\hline
\multicolumn{2}{|c|}{\textbf{Scaled velocity} $\vel$} \\ \hline
  &  \\[-0.7em]
  $[0,\, \ra]$   & $\vel(r) \, := \, \sqrt{\dfrac{\gamma(r)\, \flpr_0(r) }{\density(r)}}$,
                   using the adiabaticity. \\ \hdashline
  $r > \ra$      & constant value $\vel(r) \, = \, \atmovel \,=\,$ \num{9.8607828e-6} \si{\per\second}. 
\\ \hline
  \multicolumn{2}{|c|}{} \\[-0.6em]
  \multicolumn{2}{|c|}{Illustration of the scaled velocity $\vel$.} \\
  \multicolumn{2}{|c|}{
  \renewcommand{\datafile} {models_main.txt}
  \renewcommand{\dataA}    {cp} 
  \renewcommand{\legendA}  {$\vel$} 
  \renewcommand{\myylabel} {$\vel = c_0 / \Rsun$ (\si{\per\second})} 
  \renewcommand{\legendpos}{north east}  \hspace*{-2em}
  \setlength{\plotwidth}  {10.00cm} \setlength{\plotheight} {3.50cm}
  \pgfmathsetmacro{\scale}{1} 
\begin{tikzpicture}
\begin{axis}[width=\plotwidth, height=\plotheight,
               scale only axis, yminorticks=true,clip mode=individual,
               enlarge y limits=true,
               enlarge x limits=false,
               y label style={xshift=0cm, yshift=-0.5em},
               ylabel={\myylabel},xlabel={$r$},
               x label style={xshift=0cm, yshift=0.20cm},
               legend pos=\legendpos,
               label style     ={font=\scriptsize},
               tick label style={font=\scriptsize},
               legend style    ={font=\scriptsize\selectfont},
               legend cell align={left}
               ]
     \addplot[color=black,line width=1]  
              table[x=x,y expr=\scale*\thisrow{\dataA}]
              {\datafile}; \addlegendentry{\legendA} 

\end{axis}
\end{tikzpicture}
  } \\ \hline
\end{tabular}
\end{center} \end{table}

\begin{table}[ht!] \begin{center}
\caption{Summary of information for the $\Ehe$ and $\tEhe$.}
\label{table:ehe}
\renewcommand{\arraystretch}{1.20}
\begin{tabular}{|>{\raggedright\arraybackslash}p{.08\linewidth}|
                 >{\raggedright\arraybackslash}p{.85\linewidth}|}
\hline
\multicolumn{2}{|c|}{\textbf{$\Ehe$ and $\tEhe$}} \\ \hline
  &  \\[-0.7em]
  $[0,\, \rs]$   & $\Ehe = 0\, , \qquad\quad \tEhe=0$. \\[.5em] \hdashline
  & \\[-0.8em]
  $[\rs, \, \ra]$& $\tEhe(r) \,=\, (r - \rs)^3  \big( a_0 \, + \,    a_1 ( r - \ra) + \,    a_2 ( r - \ra)^2 \big) \, ,$ \\
                 & with $a_0=$ \num{-0.2318227}~\si{\g\per\cm\cubed\per\s\squared}, 
                        $~~~a_1=$ \num{4.1506993e4}~\si{\g\per\cm\cubed\per\s\squared}, \\[.5em]
                & \hspace*{0.4cm} $~~~a_2=$ \num{-4.8643830e9}~\si{\g\per\cm\cubed\per\s\squared}, \\[.5em]
                 & $\Ehe   = \dfrac{\tEhe(r)}{\gamma(r) \,\, \flpr_0(r)}  \, .$\\[1em] \hdashline
  & \\[-0.8em]
  $r > \ra$   & $\tEhe(r)  \,=\, \rho_0(r) \, \Phi_0(r)'\, + \, \flpr_0(r)'\, , \qquad\quad 
                 \Ehe(r)   \,=\, \dfrac{\Phi'_0(r)}{\vel(r)^2}\,\, \,-\,\dfrac{\alpha_{\flpr_{0}}(r)}{\gamma(r)}  \, .$ \\[1em]
   \hline
  \multicolumn{2}{|c|}{} \\[-0.6em]
  \multicolumn{2}{|c|}{Illustration of $\Ehe$, it is zero for $r \leq \rs$.} \\
  \multicolumn{2}{|c|}{
  \renewcommand{\datafile} {model_Ehe.txt}
  \renewcommand{\dataA}    {Ehe} 
  \renewcommand{\legendA}  {$\Ehe$} 
  \renewcommand{\myylabel} {$\Ehe$} 
  \renewcommand{\legendpos}{north east}  \hspace*{-2em}
  \setlength{\plotwidth}  {10.00cm} \setlength{\plotheight} {3.50cm}
\begin{tikzpicture}
\begin{axis}[width=\plotwidth, height=\plotheight,
               scale only axis, yminorticks=true,clip mode=individual,
               enlarge y limits=true,
               enlarge x limits=false,
               xtick={1.0,1.02,1.04},
               y label style={xshift=0cm, yshift=-0.5em},
               ylabel={\myylabel},xlabel={$r$},
               x label style={xshift=0cm, yshift=0.20cm},
               legend pos=\legendpos,
               label style     ={font=\scriptsize},
               tick label style={font=\scriptsize},
               legend style    ={font=\scriptsize\selectfont},
               legend cell align={left}
               ]
     \pgfmathsetmacro{\scale}{1} 
     \addplot[color=black,line width=1]  
              table[x=x,y expr=\scale*\thisrow{\dataA}]
              {\datafile}; \addlegendentry{\legendA} 

\end{axis}
\end{tikzpicture}
  } \\ \hline
\end{tabular}
\end{center} \end{table}
  
\begin{table}[ht!] \begin{center}
\caption{Summary of information for the scaled pressure $\flpr_0$.}
\label{table:pressure}
\renewcommand{\arraystretch}{1.20}
\begin{tabular}{|>{\raggedright\arraybackslash}p{.08\linewidth}|
                 >{\raggedright\arraybackslash}p{.85\linewidth}|}
\hline
\multicolumn{2}{|c|}{$\flpr_0$ and its derivatives} \\ \hline
  &  \\[-0.7em]
  $[0,\, \rs]$   & $\flpr_0'(r) \,=\, - \rho_0(r)\, \Phi'_0(r) \,,$ \\
                 & $\flpr_0''(r)\,=\, - \rho_0(r)\, \Phi_0''(r) \,-\, \rho_0'(r) \Phi_0'(r)\,,$ \\ 
                 & $\flpr_0(r)  \,=\,\flpr_0(\rs) + \int_r^{\rs} \rho_0(s) \Phi'_0(s) \mathrm{d}s \, ,$ \\[.1em]
                 & \qquad\qquad we have $\flpr_0(\rs) \,=\, \num{1.9519925e-19}$ \si{\g\per\second\squared\per\centi\meter\cubed}.
                 \\[.5em] \hdashline
  & \\[-0.8em]
  $[\rs, \, \ra]$& $\flpr_0'(r) \, = \, - \rho_0(r) \Phi_0'(r)\, + \,  \tEhe(r) \, ,$  \\[1em] \hdashline
  & \\[-0.8em]
  $r > \ra$   & $\flpr_0(r)  \, = \,   \flpr_0(\ra) \, e^{-\atmoalpha(r-\ra)}\, .$ \\[.1em]
                 & \qquad\qquad we have $\flpr_0(\ra) \,=\, \num{1.7392414e-19}$ \si{\g\per\second\squared\per\centi\meter\cubed}
  \\[1em]
  \hline \hline
  \multicolumn{2}{|c|}{} \\[-0.6em]
  \multicolumn{2}{|c|}{Illustration of the scaled pressure $\flpr_0$
                       and zoom on a logarithmic scale.} \\
  \multicolumn{2}{|l|}{
  \renewcommand{\datafile} {models_main.txt}
  \renewcommand{\dataA}    {pressure} 
  \renewcommand{\legendA}  {$\flpr_0$} 
  \renewcommand{\myylabel} {$\flpr_0= p_0 / \Rsun^2$ (\si{\g\per\second\squared\per\centi\meter\cubed})} 
  \renewcommand{\legendpos}{north east}
  \renewcommand{\zxmin}    {0.95}    \renewcommand{\zxminbox} {0.95} 
  \renewcommand{\zxmax}    {1.02}    \renewcommand{\zxmaxbox} {1.02} 
  \renewcommand{\zymin}    {-0.002}  \renewcommand{\zymax}    {0.002}
  \renewcommand{\xticka}   {0.97}    \renewcommand{\xtickb}   {1}
  \hspace*{-4em}
  \setlength{\plotwidth}  {7.00cm} \setlength{\plotheight} {3.50cm}
\begin{tikzpicture}
\begin{axis}[width=\plotwidth, height=\plotheight,
               scale only axis, yminorticks=true,clip mode=individual,
               enlarge y limits=true,
               enlarge x limits=false,
               y label style={xshift=0cm, yshift=-0.5em},
               ylabel={\myylabel},xlabel={$r$},
               x label style={xshift=0cm, yshift=0.20cm},
               legend pos=\legendpos,
               label style     ={font=\scriptsize},
               tick label style={font=\scriptsize},
               legend style    ={font=\scriptsize\selectfont},
               legend cell align={left}
               ]
     \pgfmathsetmacro{\scale}{1} 
     \addplot[color=black,line width=1]  
              table[x=x,y expr=\scale*\thisrow{\dataA}]
              {\datafile}; \addlegendentry{\legendA} 

     \draw[line width = 1,orange] (\zxminbox,\zymin) 
     -- (\zxmaxbox,\zymin)-- (\zxmaxbox,\zymax)-- (\zxminbox,\zymax) -- cycle ;

\end{axis}
\end{tikzpicture}
  \hspace*{-0.65cm} 
 {\raisebox{0.97cm}{\begin{tikzpicture}[]
    \coordinate (d1)   at (0,0);
    \draw[line width = 1,orange] ([xshift= 0   cm,yshift=0mm] d1) -- ([xshift=-1.cm,yshift= 0cm] d1);
  \end{tikzpicture}}}
  \setlength{\plotwidth}  {4.00cm}  \setlength{\plotheight}  {2.20cm}
  \renewcommand{\datafile} {models_main_zoom.txt}
  \hspace*{-5.5mm}{\raisebox{8mm}{
\begin{tikzpicture}[framed,background rectangle/.style={line width=1,draw=orange}]]
\begin{axis}[width=\plotwidth, height=\plotheight,
               scale only axis, yminorticks=true,clip mode=individual,
               enlarge y limits=true,
               enlarge x limits=false,
               ymode=log,
               xmin=\zxmin,xmax=\zxmax,
               xtick={\xticka, \xtickb},
               xticklabels={\num{\xticka},\num{\xtickb}},
               xlabel={},
               x label style={xshift=0cm, yshift=0.20cm},
               label style     ={font=\scriptsize},
               tick label style={font=\scriptsize},
               legend style    ={font=\scriptsize\selectfont},
               legend cell align={left}
               ]
     \pgfmathsetmacro{\scale}{1} 
     \addplot[color=black,line width=1]  
              table[x=x,y expr=\scale*\thisrow{\dataA}]
              {\datafile}; 
\end{axis}
\end{tikzpicture}}}
  } \\ \hline
\end{tabular}
\end{center} \end{table}


\section*{Acknowledgments}
  This work is supported by the Inria associated-team 
  Ants (Advanced Numerical meThods for helioSeismology)
  between project-team Inria Magique 3D and the Max 
  Planck Institute for Solar System Research in G\"ottingen.
  FF is funded by the Austrian Science Fund (FWF) 
  under the Lise Meitner fellowship M 2791-N.

\bibliographystyle{siam}
\bibliography{bibliography.bib}

\begin{thebibliography}{10}

\bibitem{barucq2018atmospheric}
{\sc H.~Barucq, J.~Chabassier, M.~Durufl{\'e}, L.~Gizon, and M.~Legu{\`e}be},
  {\em Atmospheric radiation boundary conditions for the {H}elmholtz equation},
  ESAIM: Mathematical Modelling and Numerical Analysis, 52 (2018),
  pp.~945--964.

\bibitem{barucq:hal-02423882}
{\sc H.~Barucq, F.~Faucher, D.~Fournier, L.~Gizon, and H.~Pham}, {\em {On the
  outgoing solutions and radiation boundary conditions for the vectorial wave
  equation with ideal atmosphere in helioseismology}}, Research Report RR-9335,
  {Inria Bordeaux Sud-Ouest ; Magique 3D ; Max-Planck Institute for Solar
  System Research}, Apr. 2020.

\bibitem{barucq:hal-02168467}
{\sc H.~Barucq, F.~Faucher, and H.~Pham}, {\em {Outgoing solutions to the
  scalar wave equation in helioseismology}}, Research Report RR-9280, {Inria
  Bordeaux Sud-Ouest}, August 2019.

\bibitem{OutEsaim}
\leavevmode\vrule height 2pt depth -1.6pt width 23pt, {\em Outgoing solutions
  and radiation boundary conditions for the ideal atmospheric scalar wave
  equation in helioseismology}, ESAIM: Mathematical Modelling and Numerical
  Analysis, to appear (2020).

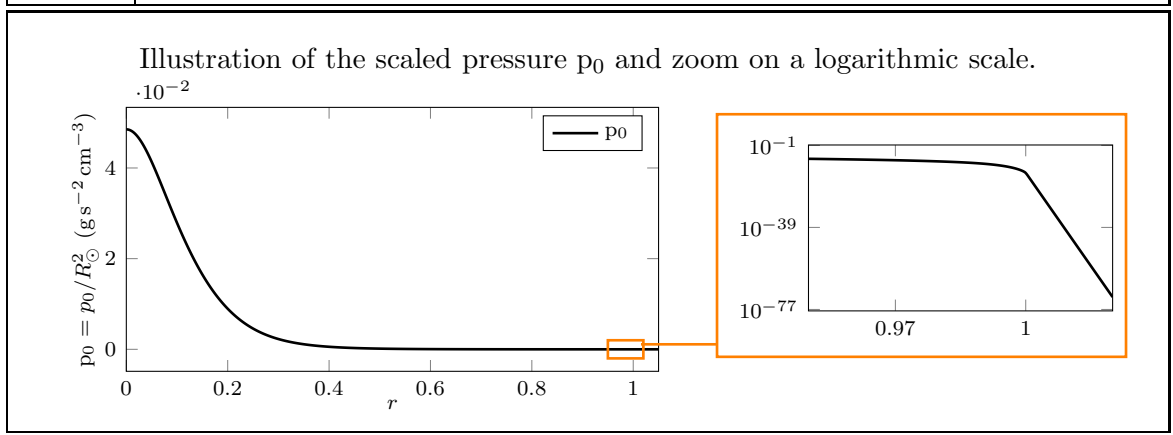
\bibitem{christensen1996current}
{\sc J.~Christensen-Dalsgaard, W.~D{\"a}ppen, S.~Ajukov, E.~Anderson, H.~Antia,
  S.~Basu, V.~Baturin, G.~Berthomieu, B.~Chaboyer, S.~Chitre, et~al.}, {\em The
  current state of solar modeling}, Science, 272 (1996), pp.~1286--1292.

\bibitem{fournier2017atmospheric}
{\sc D.~Fournier, M.~Legu{\`e}be, C.~S. Hanson, L.~Gizon, H.~Barucq,
  J.~Chabassier, and M.~Durufl{\'e}}, {\em Atmospheric-radiation boundary
  conditions for high-frequency waves in time-distance helioseismology},
  Astronomy \& Astrophysics, 608 (2017), p.~A109.

\bibitem{gizon2017computational}
{\sc L.~Gizon, H.~Barucq, M.~Durufl{\'e}, C.~S. Hanson, M.~Legu{\`e}be, A.~C.
  Birch, J.~Chabassier, D.~Fournier, T.~Hohage, and E.~Papini}, {\em
  Computational helioseismology in the frequency domain: acoustic waves in
  axisymmetric solar models with flows}, Astronomy \& Astrophysics, 600 (2017),
  p.~A35.

\bibitem{legendre2003rayonnement}
{\sc G.~Legendre}, {\em Rayonnement acoustique dans un fluide en
  {\'e}coulement: analyse math{\'e}matique et num{\'e}rique de l'{\'e}quation
  de Galbrun}, PhD thesis, 2003.

\bibitem{Piegl2012}
{\sc L.~Piegl and W.~Tiller}, {\em The NURBS book}, Springer Science \&
  Business Media, 2012.

\bibitem{NumAKress}
{\sc K.~Rainer}, {\em Numerical analysis}, Springer, 1998.

\end{thebibliography}

\end{document}